# Predicting Economic Recessions Using Machine Learning Algorithms

**Rickard Nyman[1] and Paul Ormerod[2]**

**December 2016**

Acknowledgement: we acknowledge the assistance of Oliver Rice in checking our results in Python

[1] Periander Ltd and University College London (UCL); r.nyman@periander.co.uk

[2] University College London (UCL), Volterra Partners LLP, London and Periander Ltd.; pormerod@ucl.ac.uk. Corresponding author

*Abstract*

*Even at the beginning of 2008, the economic recession of 2008/09 was not being predicted by the economic forecasting community.  The failure to predict recessions is a persistent theme in economic forecasting.  The Survey of Professional Forecasters (SPF) provides data on predictions made for the growth of total output, GDP, in the United States for one, two, three and four quarters ahead, going back to the end of the 1960s.  Over a three quarters ahead horizon, the mean prediction made for GDP growth has* never *been negative over this period. The correlation between the mean SPF three quarters ahead forecast and the data is very low, and over the most recent 25 years is not significantly different from zero.*

*Here, we show that the machine learning technique of random forests has the potential to give early warning of recessions.  We use a small set of explanatory variables from financial markets which would have been available to a forecaster at the time of making the forecast.  We train the algorithm over the 1970Q2-1990Q1 period, and make predictions one, three and six quarters ahead.  We then re-train over 1970Q2-1990Q2 and make a further set of predictions, and so on.  We did not attempt any optimisation of predictions, using only the default input parameters to the algorithm we downloaded in the package R.*

*We compare the predictions made from 1990 to the present with the actual data. One quarter ahead, the algorithm is not able to improve on the SPF predictions. Three and six quarters ahead, the correlations between actual and predicted are low, but they are very significantly different from zero.  Although the timing is slightly wrong, a serious downturn in the first half of 2009 could have been predicted six quarters ahead in late 2007.  The algorithm never predicts a recession when one did not occur.*

*We obtain even stronger results with random forest machine learning techniques in the case of the United Kingdom.*

# 1. Introduction

Over the past fifty years or so, a track record of macroeconomic forecasts and their accuracy has been built up. Economists disagree about how the economy operates, and these disagreements are reflected in, amongst other things, the specification of the relationships in macro-economic models. But, over time, no single approach has a better forecasting record than any other.

There are some general points which emerge from the literature on forecasting accuracy, especially with respect to the one year ahead predictions for growth rate of GDP. Twenty years ago, for example, a major survey article concluded that there is no real evidence to suggest that accuracy was improving over time (Fildes and Stekler, 2000). This has persisted. Indeed, there is some suggestion that accuracy has deterioriated in recent years (see for example the review of the UK Monetary Policy Committee's forecasts Stockton (2012)).

A particular problem is the very poor record of predicting recessions. The failure to forecast the financial crisis recession of the late 2000s is well known. But the same point was made in 1993 by Zarnowitz and Braun. In the United States, for example, they point out that the recessions following the 1974 and 1981 peaks in the level of output were not recognised even as they took place.

The purpose of this paper is to examine whether machine learning algorithms can improve forecasting accuracy. Varian (2014) gives a wide range of examples of the potential application of such algorithms.

Our specific focus is on short-term forecasts of real GDP growth in the United States, and in particular in whether the recession of the late 2000s could have been predicted. We also include the United Kingdom in our analysis.

Real GDP began to fall in the US on a quarter-by quarter basis in a sustained way (there was a temporary fall in 2008Q1) in 2008Q3, and growth was not resumed until 2009Q3. The previous peak level of real GDP in 2008Q2 was not reached until 2011Q2. The cumulative fall in output in the 2008/09 period was 4.1 per cent at an annualised rate, and the cumulative fall from the 2008Q2 peak until 2011Q2 was 22.9 per cent. The recession was by some margin the biggest since quarterly national accounts data began in 1947Q1. To set it in context, during the previous global financial crisis in the 1930s, real GDP fell in each year between 1930 and 1933, with a cumulative fall of 29 per cent. The 1929 peak level was not reached until 1936, making a cumulative loss of output below the previous peak of 93 per cent. So although the recession of the late 2000s was mild in comparison, it was still a serious one.

Section 2 describes the benchmarks for forecasting accuracy given by the mean predictions of the Survey of Professional Forecasters, maintained and published by the Federal Reserve Bank of Philadelphia (https://www.philadelphiafed.org/research-and-data/real-time-center/survey-of-professional-forecasters/).  Section 3 sets out the methodology we follow, section 4 describes the results for the US, section summarises the results for the UK, and section 6 offers a short conclusion.

## 2. The Survey of Professional Forecasters benchmarks

The Survey of Professional Forecasters publishes a wide range of information on economic forecasts.  It is the oldest quarterly survey of macroeconomic forecasts in the United States. The survey began in 1968 and was conducted by the American Statistical Association and the National Bureau of Economic Research. The Federal Reserve Bank of Philadelphia took over the survey in 1990.

We focus here on the short-term predictions for real GDP growth.  A complete series going back to 1968Q4 is available for one, two and three quarter ahead forecasts of GDP growth (at an annualised rate).  In other words, the SPF data in 1968Q4 for the one quarter ahead prediction is the prediction for the outturn in 1969Q1.

In terms of the one quarter ahead predictions, although the overall explanatory power is low, a regression of the actual growth rate on the mean prediction published by the SPF shows that the predictions are unbiased.  Table 1 shows the results of OLS regressions of the mean SPF prediction on each of the most recent estimate of GDP growth and the third published estimate[3].  The third estimate is released near to the end of the third month after the end of the quarter to which the data refers. We also set out in Table 1 the regressions from 1990Q2, for reasons explained in section 3 below.  We report the results with the most recent estimates of GDP growth out of interest, our comparator for the results reported in section 4 is the third estimate data.  In n-step predictions later in the paper, lagged values of variables must be used, and so to enable exact comparisons with the SPF prediction record, we start the sample in 1970Q2.

---

[3] The various vintages of GDP growth estimates are available at http://www.bea.gov/National/index.htm

**Table 1    Regressions of GDP quarter on quarter growth, annualised rate, per cent, on SPF forecasts**

Dependent variable (GDP quarter on quarter growth, annualised rate, per cent)

| Sample Period | 1970Q2 – 2016Q2 | | 1990Q2 – 2016Q2 | |
| --- | --- | --- | --- | --- |
| | Most Recent Estimate | Third Estimate | Most Recent Estimate | Third Estimate |
| | (1) | (2) | (3) | (4) |
| SPF | 0.940 | 1.054 | 1.102 | 1.175 |
| | (0.123) | (0.121) | (0.225) | (0.207) |
| Constant | 0.248 | -0.171 | -0.445 | -0.482 |
| | (0.386) | (0.379) | (0.611) | (0.563) |
| Observations | 185 | 185 | 105 | 105 |
| Adjusted R2 | 0.237 | 0.289 | 0.181 | 0.230 |
| Residual Std. Error | 2.851 | 2.799 | 2.229 | 2.051 |

The situation is quite different with the three quarter ahead predictions. For example, three quarters ahead, the mean SPF prediction has *never* been for negative growth over the entire 1970Q2 – 2016Q2 period. This reinforces the point made in section 1 above that macro forecasts are very poor in terms of their track record on recessions.

A regression of the three quarters ahead prediction for real GDP growth on the most recent estimate gives an R bar squared of 0.040, and on the third estimate of GDP of 0.050. Over the 1990Q2-2016Q2 period, they are -0.005 and 0.009 respectively.

The historical accuracy of the SPF predictions therefore provides two benchmarks for a machine learning approach:

- Given data which was available at the time a prediction would have been made, is it possible to obtain more explanatory power on one period ahead predictions
- Given data which was available at the time a prediction would have been made, is it possible to obtain *any* explanatory power on three period ahead predictions

In addition, we also examine six quarter ahead predictions out of interest, though the SPF does not publish these.

## 3. Methodology

### 3.1   Data

We try to replicate as closely as possible the situation in which a forecast would actually be made.  The dependent variable in our analysis is the third estimate of real GDP growth in the relevant quarter, rather than the most recent estimate which is now available.  The third estimate is in general the one on which policy makers would rely when trying to judge the current state of the economy at the time.

We should stress that the explanatory data set was selected on a theoretical basis, without prior investigation of how any of the variables correlated with GDP growth.  Once selected, we did not amend the data base in any way to try to improve statistical fits, a common practice in much macroeconomic time series investigations. The motivation was that there is a distinct tradition within economics of regarding recessions, especially deep ones, as being of monetary origin.  Friedman and Schwartz (1963) is a key text here, in which they argued that the Federal Reserve's monetary policies were largely to blame for the severity of the Great Depression.

We choose explanatory variables principally from financial markets, where in theory information should exist about the future state of the economy.  Perhaps more importantly in the context of trying to replicate a genuine forecasting situation, these variables are both available at the time, and their values are not subsequently revised.  We use the 3 month Treasury Bill rate, the yield on 10 year US government bonds, and the quarterly percentage change in the Standard and Poors 500 index.

We add to the set of explanatory variables the ratio of private sector debt to current price GDP, the debt data being taken from the Bank of International Settlements website.  This data usually appears with a lag of three or four months, so the most recent quarter in any forecasting situation would have to have been estimated.  Further, current price GDP is subject to revisions, but given that it is used to divide another large number of similar magnitude, any such revisions in the ratio will be very small.

### 3.2   Model estimation techniques

We compare the results of two different estimation techniques:

- Ordinary least squares regression
- Random forest machine learning

Random forests (Breiman, 2001, 2002) are machine-learning models known for their ability to cope with noisy, non-linear, high-dimensional prediction problems. Many proofs of their properties which extend the original work of Breiman are available in, for example, Biau et al. (2008) and Biau (2012).

They construct a large number of decision trees in training by sampling with replacement from the observations. Each tree in the collection is formed by first selecting at random, at each node, a small group of input coordinates to split on and, secondly, by calculating the best split based on these features in the training set. Each tree gives a prediction, and the predictions are averaged. From the point of view of the bias-variance trade-off, the ensemble of a large number of trees trained on independent bootstrap samples, each with relatively large variance but low bias, achieve much reduced variance without the introduction of additional bias.

Fernandez-Delgado et al. (2014), in a paper whose citations are rising rapidly, compare 179 classification algorithms from 17 "families" such as Bayesian, neural networks, logistic and multinomial regression, They examine their performance on 121 data sets in the University of California at Irvine machine learning repository. This repository is in standard use in machine learning research. The authors find that the random forest family of algorithms achieves the best results. The closest rival is support vector machines. There are a few of others which have good results. But the authors note that the remainder, which include Bayesian and logistic regression algorithms, "are not competitive at all". (p.3175). In an economic context, Alessi and Detken (2011) and Alessi et al. (2015) report good results with random forest algorithms in the context of early warning of banking crises.

To illustrate our approach using the example of OLS and the one quarter ahead predictions, we regress the GDP growth variable on lagged values of the explanatory variables. We used lagged values because these values would have been available to a forecaster at the time. We use lags from one through four.

Initially, we carry out the regression over the period 1970Q2 through 1990Q1, and use the resulting equation to predict 1990Q2. We then repeat the regression over the period 1970Q2 through 1990Q2, and predict 1990Q3, and so on until we regress over the period 1970Q2 through 2016Q1 and predict 2016Q2.

We save each of the predictions, and then regress the data series composed of these predictions on, in turn, the third estimates of GDP growth over the 1990Q2 through 2016Q2 period. The results are then compared to the benchmark SPF accuracy regressions reported in Table 1 above.

For three quarters ahead, we use the same sample periods, but use the third through sixth lags of the explanatory variables. In other words, we regress GDP growth at time t on the explanatory variables at times t-3, t-4, t-5 and t-6. These are the values which would have been available at the time at which a three quarter ahead prediction was made.

We also carry out the same procedure with a six quarter ahead prediction, just using the sixth and seventh lags of the explanatory variables. No direct comparison with the SPF accuracy is available, but it is of interest to see what level of accuracy could be obtained over this longer time horizon.

The same procedure is followed with the random forest algorithm. We describe the procedure above using the example of OLS simply because it is much more familiar and so less cumbersome to describe.

We use the statistical program R, and downloaded the package *randomForest* to carry out the random forest analysis[4].

We emphasise that we used the default values for the various options available for inputs in the random forest algorithm. In other words, we did not attempt in any way to optimise the accuracy of the predictions by trying different combinations of input parameters. It is very likely that better results could be obtained in this way, but this would violate the principle of replicating as far as possible an actual forecasting situation.

## 4. Results for the United States

In terms of one step ahead forecasts, neither of the two approaches gave results in terms of explanatory power which are as good as the simple regression of the third estimate of GDP growth on the mean SPF predictions, reported in Table 1 above. In general, the predictions were also biased.

The interest of the results is in the three and six step ahead predictions. As noted above, the SPF track record is very poor over the three quarter horizon, with a regression of GDP growth on the mean SPF prediction having zero explanatory power over the 1990 Q2-2016 Q2 period.

---

[4] Oliver Rice kindly used the RandomForestRegressor tool in the scikit-learn package to check the validity of our results. The same parameters as the defaults in the R package were used as far as possible

The regression of GDP growth on the predictions from the linear model gave an R bar-squared of only 0.009, not significantly different from zero. The random forest approach gives a considerably higher R bar squared of 0.149.

Over the six period ahead horizon, the linear model has an R bar-squared of 0.025, significantly different from zero at a p-value of 0.059. Although in scientific terms the explanatory power of the random forest approach remains low with an R bar-squared of 0.091, it is significantly different from zero at a p-value of 0.001.

We report the regression of the actual third estimate of GDP growth on the random forest predictions over the three and six quarter horizons in Table 2 below

**Table 2  Regressions of GDP quarter on quarter growth, annualised rate, per cent, third estimate on random forest predictions made three and six quarters previously**

Dependent variable:  GDP quarter on quarter growth, annualised rate, per cent, third estimate

|  | Three quarters ahead | Six quarters ahead |
|---|---|---|
| Prediction | 0.674 | 0.548 |
|  | (0.154) | (0.162) |
| Constant | 0.570 | 1.105 |
|  | (0.488) | (0.466) |
| Observations | 105 | 105 |
| Adjusted R2 | 0.149 | 0.091 |
| Residual Std. Error | 2.157 | 2.229 |

The predictions are in each case biased, the coefficients on the predictions being significantly different from one, and in the case of the six quarter ahead predictions the constant term is significantly different from zero. However, they each have explanatory power which is significantly different from zero. In the case of the three quarter ahead, this compares with the zero explanatory power of the mean SPF forecasts, the SPF not publishing a track record on six quarter ahead prediction.

Figure 1 plots the actual third estimate GDP growth and the predictions made by the random forest approach six quarters previously, over the 1990Q2-2016Q2 period.

**Figure 1    Actual annualised quarter on quarter third estimate US GDP growth, per cent, and random forest predictions made six quarters previously, 1990Q2 – 2016Q2**

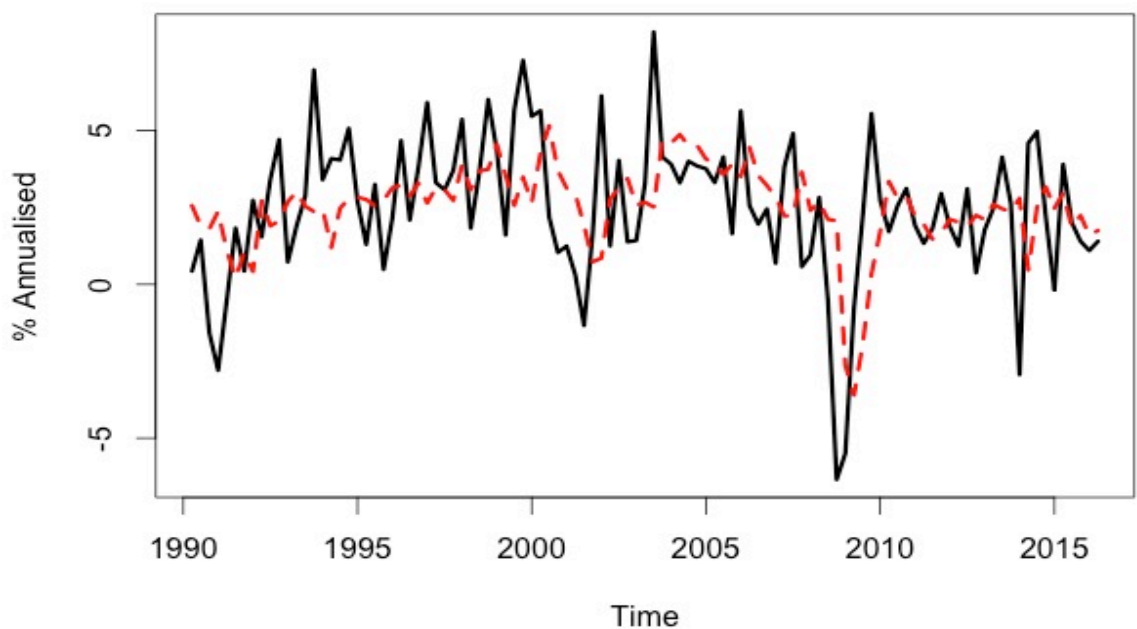

**Note:** *solid black line is actual, dotted red line predicted*

Intriguingly, although the random forest predictions would not have got the exact timing of the recession in the winter of 2008/09 correct, a serious recession would have been predicted for early 2009 eighteen months previously.

| Period | Actual annualised third estimate quarter on quarter real GDP growth | Random forest predictions made six quarters previously |
|---|---|---|
| 2008Q1 | 0.96 | 2.40 |
| 2008Q2 | 2.83 | 2.69 |
| 2008Q3 | -0.51 | 2.11 |
| 2008Q4 | -6.34 | 2.07 |
| 2009Q1 | -5.49 | -2.63 |
| 2009Q2 | -0.74 | -3.60 |
| 2009Q3 | 2.24 | -1.93 |
| 2009Q4 | 5.55 | 0.33 |

The results are really quite striking. For example, in 2007Q3, a prediction of a negative growth rate of -2.63 per cent in 2009Q1 could have been made. The overall depth of the recession predicted for 2009Q1-2009Q3 was obviously not as big as the actual recession itself, but it is very similar to the recession of the early 1980s, which until 2008/09 was distinctly the largest recession experienced since the 1930s.

There are two other periods of recession in the period on which we focus: the winter of 1990/91 and 2001 (technically, there was only one period of negative growth in the third estimate data for 2001, but growth was close to zero in other quarters). In both cases, the random forest approach predicts six quarters ahead a marked slowdown in growth, but again slightly later than occurred. The approach does not predict a recession, but several periods of growth under 1 per cent (at an annualised rate). In general, periods with growth at this low level are associated, for example, with rising unemployment.

The random forest does *not* predict a recession in periods when one did not in fact take place.

## 5. Results for the United Kingdom

We describe these results considerably more briefly, adopting the same approach as above. There is no historical record of actual forecasts made which is similar to that of the SPF in the US, so we are unable to make this direct comparison. But, in general, the macroeconomic forecasting record is similar in the two countries.

As explanatory variables, we use the 3 month Treasury Bill rate, the yield on 10 year UK government bonds, the quarterly percentage change in the FTSE All Share Index and the ratio of private sector debt to current price GDP. The Bank of England statistic database, our main source, does not go back quite as far as is the case with the US sources, so we predict over the period 1995Q1 through 2016Q1, using the Office for National Statistics second estimate of real GDP.

Looking three quarters ahead, the linear regression using four relevant lags of the explanatory variables gives an R bar squared for the regression of the actual second estimate of GDP growth on its predicted value of 0.004. In contrast the R bar squared using the predictions from the random forest model is 0.246.

Looking six quarters ahead, the linear regression using four relevant lags of the explanatory variables gives an R bar squared for the regression of the actual second estimate of GDP growth on its predicted value of 0.042. In contrast the R bar squared using the predictions from the random forest model is 0.290.

Figure 2 plots the actual and the six quarter ahead prediction.

**Figure 2    Actual annualised quarter on quarter third estimate UK GDP growth, per cent, and random forest predictions made six quarters previously, 1995Q1 – 2016Q1**

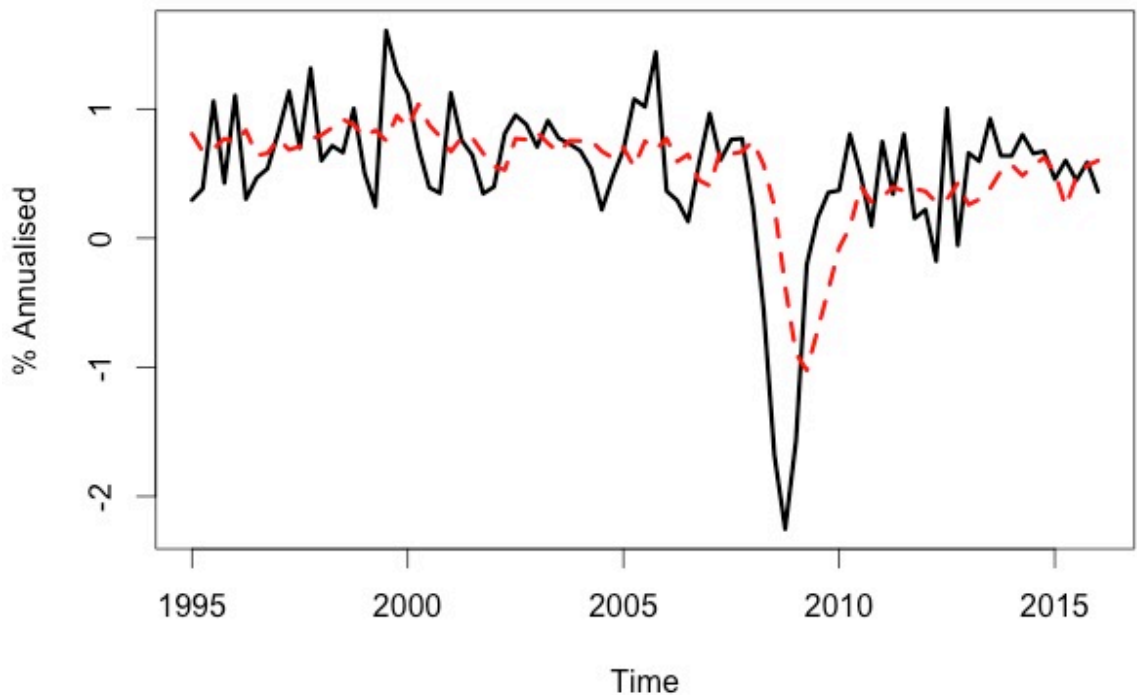

**Note:**  *solid black line is actual, dotted red line predicted*

## 6.  Concluding remarks

We have tried, as far as it is possible, to replicate an actual forecasting situation starting for the United States in 1990Q2 and moving forward a quarter at a time through to 2016. We use a small number of lags on a small number of financial variables in order to make predictions.

In terms of one step ahead predictions of real GDP growth, we have not been able to improve upon the mean forecasts made by the Society of Professional Forecasters.

However, even just three quarters ahead, the SPF track record is very poor. A regression of actual GDP growth on the mean prediction made three quarters previously has zero explanatory power, and the SPF predictions never indicated a single quarter of negative growth. The random forest approach improves very considerably on this.

Even more strikingly, over a six period ahead horizon, the random forest approach would have predicted, during the winter of 2007/08, a severe recession in the United States during 2009, ending in 2009Q4.

Again to emphasise, we have not attempted in any way to optimise these results in an *ex post* manner. We use only the default values of the input parameters into the machine learning algorithm, and use only a small number of explanatory variables.

We obtain qualitatively similar results for the UK, though the predictive power of the random forest algorithm is even better than it is for the United States.

As Ormerod and Mounfield (2000) show, using modern signal processing techniques, the time series GDP growth data is dominated by noise rather than by signal. So there is almost certainly a quite restrictive upper bound on the degree of accuracy of prediction which can be achieved. However, machine learning techniques do seem to have considerable promise in extending useful forecasting horizons and providing better information to policy makers over such horizons.

**References**

Alessi, L. and Detken, C., 2011. Quasi real time early warning indicators for costly asset price boom/bust cycles: A role for global liquidity. *European Journal of Political Economy*, *27*(3), pp.520-533

Alessi, L., Antunes, A., Babecký, J., Baltussen, S., Behn, M., Bonfim, D., Bush, O., Detken, C., Frost, J., Guimaraes, R. and Havranek, T., 2015. Comparing different early warning systems: Results from a horse race competition among members of the macro-prudential research network.

Biau, G., 2012. Analysis of a random forests model. *Journal of Machine Learning Research*, *13*(Apr), pp.1063-1095.

Biau, G., Devroye, L. and Lugosi, G., 2008. Consistency of random forests and other averaging classifiers. *Journal of Machine Learning Research*, *9*(Sep), pp.2015-2033.

Breiman, L., 2001. Random forests. *Machine learning*, *45*(1), pp.5-32

Breiman, L., 2002. Manual on setting up, using, and understanding random forests v3. 1. *Statistics Department University of California Berkeley, CA, USA*

Fildes, R. and Stekler, H., 2002. The state of macroeconomic forecasting. *Journal of Macroeconomics*, *24*(4), pp.435-468

Ormerod, P. and Mounfield, C., 2000. Random matrix theory and the failure of macro-economic forecasts. *Physica A: Statistical Mechanics and its Applications*, *280*(3), pp.497-504.

Stockton, D., 2012. *Review of the Monetary Policy Committee's Forecasting Capability,* presented to the Court of the Bank of England http://www.bankofengland.co.uk/publications/Documents/news/2012/cr3stockton.pdf

Varian, H.R., 2014. Big data: New tricks for econometrics. *The Journal of Economic Perspectives*, *28*(2), pp.3-27

Zarnowitz, V. and Braun, P., 1993. Twenty-two years of the NBER-ASA quarterly economic outlook surveys: aspects and comparisons of forecasting performance. In *Business cycles, Indicators and Forecasting* (pp. 11-94). University of Chicago Press.